# Optimized 2D CA-CFAR for Drone-Mounted Radar Signal Processing Using Integral Image Algorithm

Budiman P.A. Rohman*, Muhammad Bagus Andra, and Masahiko Nishimoto
Graduate School of Science and Technology, Kumamoto University

*Abstract*- Buried survivor detection in the post-disaster environment by employing radar as sensor is an appealing approach. However, the implementation in the real field is challenging especially for large observation missions. Mounting the radar on the flying drone is the most promising solution. In this case, since the limitations of drones such as low computer specification and limited power resources, an efficient radar data processing is crucially required. Hence, this paper study about the implementation of the integral image technique to optimize the computation of the signal processing step of ultra-wideband impulse radar signatures. The evaluation was held on the single board computer mounted on the developed multisensory drone. The results confirm that the developed method can relatively reduce the data processing time.

*Index Terms*- vital sign, human respiration, target detection, radar, signal processing.

## I. INTRODUCTION

BURIED survivor detection in the post-disaster environment became an important issue in the disaster management. Therefore, the search and rescue mission is very important in this period. The use of radar technology for detecting buried survivor interests many researchers [1]. For a large observation area, the use of drone-mounted radar is a possible technique.

In our previous study, we have developed a multisensory drone for disaster search and rescue mission. This drone employed an ultra-wideband (UWB) impulse radar, microphone array, camera, and medium-range RFID reader [1][2]. The radar sensor in this drone aims to detect the survivor trapped by the damaged building debris through an advanced signal processing technique. However, since the drone has some limitations especially in the low computer specification and power resources, the processing should be efficient. Thus, a robust and fast signal processing technique is needed.

Inspired by the success of the implementation of integral image algorithm [3] in some studies such as in coastal surveillance FMCW radar [4] and GPU based pulsed Doppler signal processor [5], this paper discusses the implementation of the algorithm on ultra-wideband radar signal processing technique in the case of buried survivor detection. The integral image will be implemented on the single-board computer (SBC) installed on the drone. By this study, the advantage of the implementation of the techniques can be exposed.

## II. SIGNAL PROCESSING TECHNIQUE

### A. Primary signal processing technique

The radar works by recording the reflection in a certain period. Assuming that $X \in R^{M \times N}$ is the two-dimensional signal recorded by radar where $M$ and $N$ corresponding with the fast time and slow time domain, respectively. Simply, the fast time-domain reflects the object distance while the slow time domain brings the respiration information including its periodic pattern.

The signal processing in this study implements the primary signal processing steps which are commonly used in the vital signal detection research published in [6]. Firstly, the signal recorded from the radar in the fast time domain is pre-processed using DC-shift removal and linear trend subtraction. Then, in the slow time domain, the signal is filtered by the mean filter which is then normalized. After that, each signal in the slow time domain is transformed into the frequency domain by Fast Fourier Transform to extract the human vital sign. In this study, we only consider the respiration signal with a frequency range of 0.3-0.8 Hz. From this step, we apply the integral image technique which is then processed by two-dimensional cell averaging constant false alarm rate (CA-CFAR) processing.

### B. CA-CFAR and Integral Image

CA-CFAR is a simple and popular thresholding method in radar technology. CA-CFAR assumes the binary condition $H_1$ and $H_0$ indicating the presence and absence of the target in the cell under test (CUT) expressed by

$$CUT = \begin{cases} H_0 \text{ if } CUT < T \\ H_1 \text{ if } CUT > T \end{cases} \quad (1)$$

The false alarm rate is determined beforehand to compute a threshold, with the average value of several adjacent cells surround the CUT used as a reference. The threshold $T$ can be written as

$$T = \alpha A \quad \text{where} \quad \alpha = N(P_{fa}^{1/N} - 1) \quad (2)$$

where $\alpha$ is a constant value, $P_{fa}$ a predetermined value of the false alarm rate, $A$ the average value of $N$ neighbor cells, and $N$ the selected number of cells surrounding CUT. The guard cell is also needed to remove the effect of target spectral leakage.

The integral image or summed area table technique is an algorithm for efficiently generating the sum of values in a rectangular subset of a grid. In this image/table, if we go to any point $(x, y)$ then at this table entry we will come across a value. This value s is the sum of all the pixel values above, to the left and of course including the original pixel value i of $(x, y)$ itself as equation below (see Fig.1):

$$s(x,y) = i(x,y) + s(x-1,y) + s(x,y-1) - s(x-1,y-1) \quad (3)$$

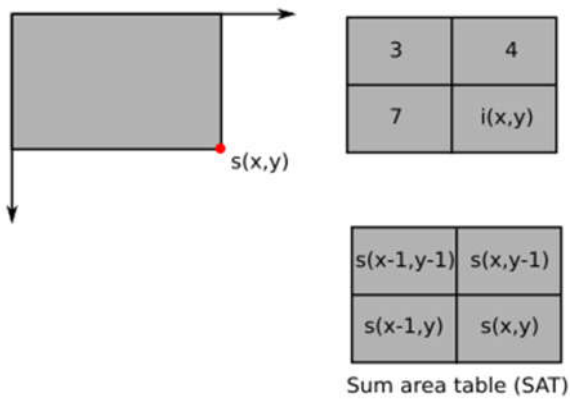

Fig. 1. Concept of integral image algorithm

Then, to calculate the sum of the region on the image we just take four points ($L_1$, $L_2$, $L_3$, and $L_4$) from the summed-area table (see Fig.2) and calculate as

$$A = L_1 + L_4 - (L_2 + L_3) \qquad (4)$$

In this study, this efficient processing of the sum calculation is taken to compute the two-dimensional CA-CFAR. Commonly, the process is held repetitively so that by using the integral image, the process will be faster.

## III. EXPERIMENTAL STUDY AND RESULT

As radar, we use Cayenne UWB impulse radar. The pulse transmitted is monocycle pulse with bandwidth 1.5-6.0GHz. The sampling rate of raw data obtained is around 39GS/s. Each of the data consists of 600 traces where each trace has 1024 sample points. In the next, the traces are named as slow time dimension and sample as fast time domain. As a barrier, we use a brick wall. We use a developed artificial vital sign actuated by microcontroller as the target. This system has displacement within 1 cm and the frequency is exactly constant 0.3 Hz.

In the development and evaluation of computation time, we use a workstation with specification: Intel Core i5-6200U processor, 8192 MB RAM. We also test the algorithm in the SBC installed on the developed drone. The computer is Raspberry-Pi 3B featured with a quad-core 64-bit ARM Cortex A53 clocked at 1.2 GH and 1 GB LPDDR2-900 RAM.

Table I shows the comparison of CA-CFAR computation time with and without the integral image. It can be seen clearly that in all CA-CFAR configurations, the use of integral image can reduce the computation of the thresholding process up to 12 times. However, compared with the overall signal processing of detection as shown by Table II, this optimization is quite small.

## IV. CONCLUSION

This paper discusses the implementation of the integral image algorithm on the radar signal processing in the case of buried survivor detection in the post-disaster area. The method works on the low-cost and compact single board computer mounted on the developed multisensory drone. We apply the primary signal processing techniques to extract the vital sign based on FFT and 2D CA-CFAR. The evaluation results reveal that the implementation of integral image on the threshold processing can reduce the computation time of detection.

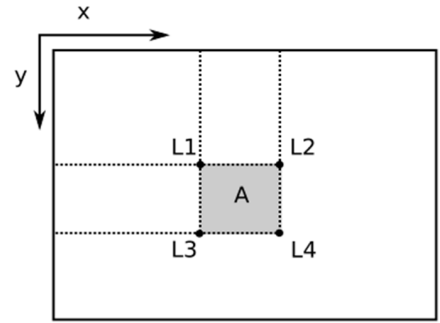

Fig. 2. Step of calculation of region on the image

TABLE I.
COMPUTATION TIME USING CA-CFAR AND II CA-CFAR (ON WORKSTATION)

| CFAR Parameters | | Computation Time (s) | | Ratio |
|---|---|---|---|---|
| Guard Cell | Background Cell | CA-CFAR | II CA-CFAR | |
| 4 | 8 | 0.126 | 0.026 | 4.8:1 |
| 4 | 12 | 0.201 | 0.016 | 12.6:1 |
| 8 | 8 | 0.175 | 0.015 | 11.7:1 |
| 8 | 12 | 0.184 | 0.019 | 9.7:1 |

TABLE II.
COMPUTATION TIME OF EACH SIGNAL PROCESSING STEP (ON SBC)

| Signal Processing Step | Time-consuming (s) | |
|---|---|---|
| | CA-CFAR | II CA-CFAR |
| Signal Recording | 8.74±0.00 | |
| Signal Processing I, Fast Time | 12.29±0.02 | |
| Signal Processing II, Slow Time | 16.37±0.02 | |
| Signal Threshold, 2D CA-CFAR | 6.62±0.00 | 1.17±0.00 |
| Total | **44.02±0.04** | **38.57±0.02** |

## V. REFERENCES

[1] B.P.A. Rohman, M.B. Andra, H.F. Putra, D.H. Fandiantoro, and M. Nishimoto. "Multisensory surveillance drone for survivor detection and geolocalization in complex post-disaster environment." in Proc. 2019 IEEE International Geoscience and Remote Sensing Symposium.
[2] M.B. Andra, B.P.A. Rohman, and T. Usagawa. "Feasibility evaluation for keyword spotting system using mini microphone array on uav." In Proc. 2019 IEEE International Geoscience and Remote Sensing Symposium.
[3] P. Viola and M. Jones. "Robust real-time object detection". *International Journal of Computer Vision* (2002) pp. 1-24
[4] T. Miftahushudur, D. Kurniawan, and B.P.A. Rohman. "Summed area table for optimizing of processing time on CA CFAR algorithm." in Proc. 2015 International Conference on Radar, Antenna, Microwave, Electronics and Telecommunications.
[5] C.J. Venter, H. Grobler, and K. A. AlMalki. "Implementation of the CA-CFAR algorithm for pulsed-Doppler radar on a GPU architecture." In Proc. 2011 IEEE Jordan Conference on Applied Electrical Engineering and Computing Technologies.
[6] Y. Xu, et al., "A novel method for automatic detection of trapped victims by ultrawideband radar.", *IEEE Transactions on Geoscience and Remote Sensing*, vol.50, no.8, pp.3132-3142, 2012.